\pdfoutput=1

\documentclass[
aps,
prl,
superscriptaddress,
twocolumn,
twoside,
floatfix,
nofootinbib,
a4paper
]{revtex4-2}

\usepackage[utf8]{inputenc}
\usepackage[T1]{fontenc}
\usepackage[british]{babel}

\usepackage[sc,osf]{mathpazo}
\usepackage[babel]{microtype}

\usepackage{
amsmath,
amssymb,
amsthm,
amsfonts,
bm,
mathrsfs,
mathtools,
bbm
}

\usepackage{braket}
\usepackage{centernot}

\usepackage{graphicx}
\usepackage{subcaption}
\usepackage{float}
\usepackage{booktabs}
\usepackage{tabularx}
\usepackage{multirow}
\usepackage{array}
\usepackage{bigstrut}
\usepackage{makecell}

\usepackage[table,dvipsnames]{xcolor}

\usepackage{pgf,tikz}

\usepackage{algorithm}
\usepackage{algorithmic}

\usepackage{enumitem}
\usepackage{comment}
\usepackage{xspace}
\usepackage{ragged2e}

\DeclareCaptionJustification{justified}{\justifying}
\usepackage[
colorlinks=true,
linkcolor=Blue,
citecolor=Magenta,
urlcolor=Blue
]{hyperref}

\newcommand{\be}{\begin{equation}}
\newcommand{\ee}{\end{equation}}

\newcommand{\bea}{\begin{eqnarray}}
\newcommand{\eea}{\end{eqnarray}}

\newcommand{\tr}{\operatorname{Tr}}

\newcommand{\I}{\mathbf{1}}

\newcommand{\la}{\langle}
\newcommand{\ra}{\rangle}

\makeatletter
\newtheorem*{rep@theorem}{\rep@title}

\newcommand{\newreptheorem}[2]{%
\newenvironment{rep#1}[1]{%
\def\rep@title{#2 \ref{##1}}%
\begin{rep@theorem}}
{\end{rep@theorem}}}
\makeatother

\newtheorem{thm}{Theorem}
\newreptheorem{thm}{Theorem}

\begin{document}


\title{Limits on Broadcasting Genuine Multipartite Entanglement in Quantum Networks}

\author{Pritam Roy}
\affiliation{S. N. Bose National Centre for Basic Sciences, Block JD, Sector III, Salt Lake, Kolkata 700106, India}

\author{William John Munro}
\affiliation{Okinawa Institute of Science and Technology Graduate University, Okinawa 904-0495, Japan}

\author{Shashank Gupta}
\email{shashankg687@gmail.com}
\affiliation{Indian Institute of Technology Indore, Khandwa Road, Simrol, Indore 453552, Madhya Pradesh, India}

\begin{abstract}
We establish operational limits on the broadcasting of genuine multipartite entanglement (GME) in quantum networks. Using a distributed protocol in which each of N parties locally implements optimal 1 $\to$ 2 cloning via beam-splitter interactions, we derive exact expressions for the broadcast fidelity of Greenberger–Horne–Zeilinger (GHZ), W states, and Cluster states in a limited setting. We show that the fidelity decays exponentially with system size as [c(R)]$^N$, providing a quantitative expression of multipartite entanglement monogamy in the broadcasting setting, and that both state families share a universal normalisation factor arising from independent post-selection probabilities. Most significantly, we prove a no-go result for simultaneous GME certification: for all reflectivities and all system sizes, the two broadcast copies cannot be simultaneously certified as genuinely multipartite entangled within the standard framework of fidelity-based witnesses. We further check this behaviour for three- and four-party cluster states, finding consistent results that support the generality of the no-go beyond the GHZ and W families. This obstruction arises from the redistribution of multipartite coherence, which both reduces the achievable fidelity and increases the corresponding certification threshold. Our results reveal a fundamental trade-off between the broadcastability of multipartite entanglement and its operational certifiability, and delineate intrinsic limits on entanglement distribution in quantum networks.
\end{abstract}


\maketitle


\textit{Introduction.---} The quantum no-cloning theorem~\cite{quantum_cloning_review_2005} forbids the perfect copying of an arbitrary quantum state, placing a fundamental constraint on the manipulation of quantum information. In practical settings, however, approximate cloning remains possible~\cite{optimal_cloning_1997,Kay2016,high_dimensional_cloning_2017}, and a substantial body of work has explored the extent to which quantum states---and more generally, quantum correlations---can be distributed across multiple systems. In particular, it has been shown that bipartite entanglement can be broadcast into two imperfect copies using local operations~\cite{cloning_entanglement_2020}, establishing that entanglement itself may be approximately replicated. A natural question is whether this capability extends to genuine multipartite entanglement (GME)---the strongest form of multiparty quantum correlation. GME is a key resource for quantum networks, underpinning tasks such as multiparty secret sharing~\cite{enhanced_secret_sharing_2024}, measurement-based quantum computation, and distributed quantum communication~\cite{anonymous_transmission_2018}. In this setting, the ability to broadcast GME from a single source to multiple nodes would be highly desirable. At the same time, multipartite entanglement is subject to stronger monogamy constraints than its bipartite counterpart~\cite{ckw_monogamy}, suggesting the possibility of more restrictive limitations on its distribution. This raises a central question: to what extent can genuine multipartite entanglement be broadcast across a quantum network while remaining operationally certifiable?

In this Letter, we address this question by analysing the broadcasting of GME under a physically natural class of distributed protocols, in which each party locally implements an optimal $1 \to 2$ cloning operation~\cite{cloning_entanglement_2020,buzek_hillery_1996}. We obtain three main results. First, the broadcast fidelity decays exponentially with the number of parties, $\mathcal{F} \sim [c(R)]^N$, providing a quantitative expression of entanglement monogamy in the multipartite broadcasting setting~\cite{ckw_monogamy}. Second, both GHZ and W state families---representing the two canonical benchmark families (which exhaust the genuine tripartite SLOCC classes for three qubits, though additional classes exist for $N > 3$)
~\cite{dur_three_qubits_2000}---share a universal normalisation denominator, which we identify as the product of independent single-site post-selection probabilities. Third, and most significantly, we establish a no-go result for simultaneous GME certification: for both state families, there exists no choice of reflectivity for which both broadcast copies can be simultaneously certified as genuinely multipartite entangled within the framework of fidelity-based witnesses~\cite{guhne_toolbox_2007,guhne_toth_2009,Haffner2005Scalable}. This result highlights a fundamental tension between the distribution of multipartite entanglement and its operational detection. While approximate copies can be generated, the broadcasting process redistributes multipartite coherence in a manner that both suppresses fidelity and raises the corresponding certification threshold. As a consequence, multipartite entanglement cannot be simultaneously replicated and certified within this operational framework, delineating intrinsic limits on entanglement sharing in quantum networks. We emphasise that these results apply to the specific protocol of local independent cloning via beam splitters and certification by fidelity-based witnesses. Collective operations across parties, or alternative witness constructions, could in principle circumvent these limitations and remain important open questions.



\textit{GME and its certification.---} Consider an $N$-partite quantum system. A pure state $\ket{\psi}$ is genuinely multipartite entangled if it is entangled across every bipartition. For three qubits, D\"ur, Vidal, and Cirac \cite{dur_three_qubits_2000} showed that there are exactly two inequivalent SLOCC classes of genuine tripartite entanglement: the GHZ and W families. For general $N > 3$, the number of SLOCC classes grows, but GHZ and W states remain two canonical and widely studied benchmark families. The GHZ state is,
\begin{equation}
    \ket{\text{GHZ}_N(\alpha)} = \alpha\ket{0}^{\otimes N} + \sqrt{1-\alpha^2}\ket{1}^{\otimes N},
    \label{eq:GHZ}
\end{equation}
with maximal entanglement at $\alpha = 1/\sqrt{2}$; and the W state,
\begin{equation}
    \ket{W_N} = \frac{1}{\sqrt{N}}\big(\ket{10\cdots0} + \ket{01\cdots0} + \cdots + \ket{00\cdots1}\big).
    \label{eq:W}
\end{equation}

To certify GME, we employ fidelity-based entanglement witnesses, which are widely used and experimentally natural \cite{guhne_toolbox_2007,guhne_toth_2009}. For any target GME state $\ket{\phi}$, the witness $\mathcal{W} = \beta\,\I - \ket{\phi}\bra{\phi}$ certifies a state $\rho$ as GME whenever $\mathcal{F}(\rho, \ket{\phi}) > \beta$, where $\beta = \max_{\ket{\psi}\in\text{bisep}} |\langle\psi|\phi\rangle|^2$ is the maximum squared overlap with any biseparable state.

For the \emph{maximally} entangled GHZ state ($\alpha = 1/\sqrt{2}$), $\beta = 1/2$ \cite{guhne_toolbox_2007, guhne_toth_2009}. For a \emph{non-maximally} entangled GHZ state, the threshold is state-dependent:
\begin{equation}
    \beta(\alpha) = \max\!\big(\alpha^2,\; 1-\alpha^2\big),
    \label{eq:beta_alpha}
\end{equation}
which exceeds $1/2$ for any $\alpha \neq 1/\sqrt{2}$, making certification strictly more demanding. For the W state, $\beta_W = (N-1)/N$ ~\cite{guhne_toolbox_2007, guhne_toth_2009,Haffner2005Scalable}. We stress that fidelity witnesses provide a sufficient (not necessary) condition for GME: failure of this witness does not exclude GME detected by other means, but fidelity witnesses are the standard operational tool in current experiments.


\textit{Broadcasting protocol.---} We carefully distinguish three related concepts: (i)~the \emph{physical beam splitter}, which unitarily mixes two bosonic modes; (ii)~the \emph{post-selected qubit map}, obtained by projecting the output back onto the single-photon subspace; and (iii)~the \emph{optimal $1 \to 2$ cloning transformation}, to which the post-selected map is equivalent.

\begin{figure}[h!]
	\includegraphics[width=0.4\textwidth]{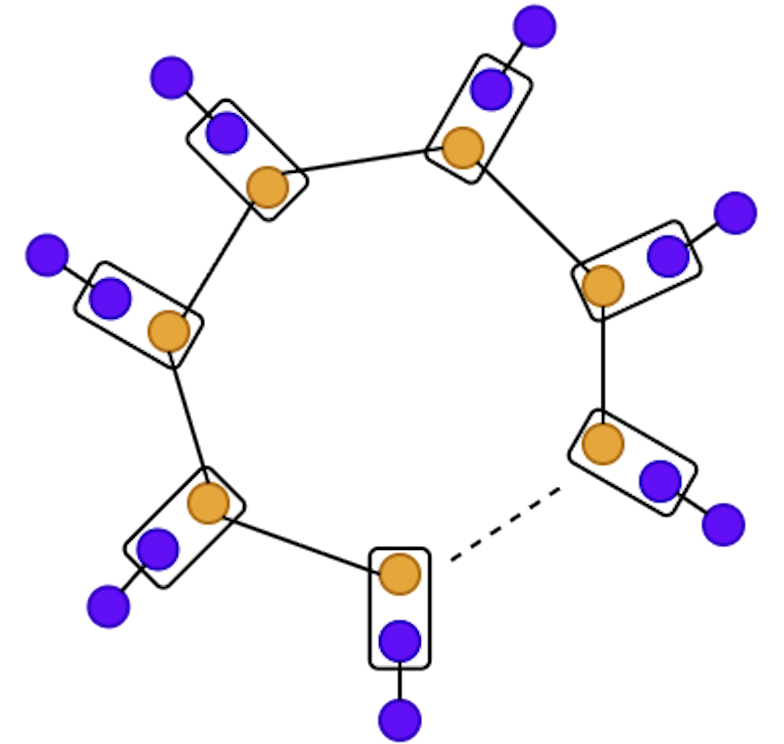}
	\caption{\footnotesize{$N$-party GME broadcasting protocol. Each party's input qubit interacts with one ancilla qubit at a beam splitter (BS) of reflectivity $R$. The product map $\hat{U}_{\text{broad}} = \prod_k \hat{U}_k^{\text{eff}}$ yields two approximate $N$-partite copies in the transmitted and reflected modes. The factorised structure means each beam splitter contributes an independent degradation factor.}}
	\label{fig:protocol}
\end{figure}

Each of $N$ parties possesses one qubit of the GME state $\ket{\Phi_N}$ and is provided with an ancillary pair $\ket{\chi_j}_{a_j b_j} = \gamma_j \ket{01} - \sqrt{1 - \gamma_j^2}\,\ket{10}$ (with $\gamma_j = 1/\sqrt{2}$ for symmetric broadcasting). The full initial state is $\ket{\Psi_{\text{in}}} = \ket{\Phi_N} \otimes \bigotimes_{j=1}^{N} \ket{\chi_j}_{a_j b_j}$, spanning a $3N$-qubit Hilbert space.
Now each party sends their input qubit and one ancilla qubit through a beam splitter. In terms of the bosonic creation operators, the standard unitary beam-splitter transformation is $\hat{U}_{\text{BS}} = \exp[\theta(\hat{a}^\dagger \hat{b} - \hat{a}\hat{b}^\dagger)]$, where $\cos^2\theta = T$ and $\sin^2\theta = R$.
Restricting to the single-photon subspace at each output port, the effective transformation (with success probability as $T^2+R^2$) takes the form of a partial swap:
\begin{equation}
    \hat{U}_k^{\text{eff}} = T\,\hat{\I}_{i,j} - R\,\hat{S}_{i,j},
    \label{eq:Ueff}
\end{equation}
where $\hat{S}_{i,j}$ is the swap operator exchanging the input qubit ($i$) and ancilla qubit ($j$) of party $k$, and $T = 1-R$ parametrises the transmissivity within the post-selected subspace. We note that while the full bosonic beam splitter is unitary with $T + R = 1$, the post-selected qubit map is successful with probability $T^2+R^2<1$; the two parametrisations are related by the post-selection projection (see Supplemental Material~\cite{supplemental} for the full derivation).
This post-selected map is identical to the Bu\v{z}ek--Hillery optimal universal $1 \to 2$ qubit cloner \cite{buzek_hillery_1996,cloning_entanglement_2020}. Because the $N$ parties operate independently---as is natural in a distributed quantum network where they are spatially separated---the full broadcasting map factorises:
\begin{equation}
    \hat{U}_{\text{broad}} = \prod_{k=1}^{N} \hat{U}_k^{\text{eff}}.
    \label{eq:Ubroad}
\end{equation}
This product structure is the key origin of the multiplicative fidelity scaling discussed below. After the interaction, two approximate $N$-partite copies are extracted: Copy~1 (transmitted modes) and Copy~2 (non-interacting ancilla modes), with post-selection on single-photon occupation at each output port (see Fig.~\ref{fig:protocol} and Supplemental Material~\cite{supplemental} for details).

\begin{thm}\label{thm:GHZ}
(GHZ broadcasting) For the $N$-partite GHZ state $\ket{\textup{GHZ}_N(\alpha)}$ undergoing local beam-splitter broadcasting, the transmitted copy fidelity is
\begin{equation}
    \mathcal{F}_t^{(N)}(\alpha, R) = \frac{P_t^{(N)}(\alpha, R)}{\big(R^2 - R + \tfrac{1}{3}\big)^N},
    \label{eq:fidelity_GHZ_N}
\end{equation}
where $P_t^{(N)}(\alpha, R)$ is a polynomial of degree $2N$ in $R$ (see Supplemental Material~\cite{supplemental}  for the closed-form expression). The fidelity decays exponentially:
\begin{equation}
     \mathcal{F}_t^{(N)}(R) \sim \big[c_{\textup{GHZ}}(R)\big]^N, \quad N \to \infty,
     \label{eq:asymptotic_GHZ}
\end{equation}
with $c_{\textup{GHZ}}(0) = 1$ and $c_{\textup{GHZ}}(1/2) = 1/2$.
\end{thm}

\begin{thm}\label{thm:W}
(W broadcasting) The symmetric $N$-partite W state has the same fidelity structure with an identical denominator:
\begin{equation}
    \mathcal{F}_t^{(N)}(R) = \frac{Q_t^{(N)}(R)}{\big(R^2 - R + \tfrac{1}{3}\big)^N},
    \label{eq:fidelity_W_N}
\end{equation}
where $Q_t^{(N)}$ involves a weighted sum over the $N$ single-excitation basis states with beam-splitter redistribution coefficients (see Supplemental Material~\cite{supplemental}  Sec.~6). The exponential decay $\mathcal{F}_t^{(N)} \sim [c_W(R)]^N$ satisfies $c_W(R) \geq c_{\textup{GHZ}}(R)$ for all $R \in [0, 1/2]$.
\end{thm}

The exponential decay is a direct consequence of the factorised structure of the local cloning maps (Eq.~\eqref{eq:Ubroad}): each beam splitter contributes an independent multiplicative factor to the fidelity, and the full fidelity is the product of $N$ such factors. This behaviour is physically expected from entanglement monogamy \cite{ckw_monogamy}---the original $N$-partite entanglement is a finite resource shared between the two copies---though we note that the exponential scaling follows from the product structure of the protocol rather than constituting an independent monogamy bound.

The inequality $c_W \geq c_{\text{GHZ}}$ reveals that the decay rate is sensitive to the entanglement structure, not just the local mixing. Physically, the GHZ state concentrates its weight on the extremal Fock layers $\ket{0}^{\otimes N}$ and $\ket{1}^{\otimes N}$, which are maximally affected by the beam splitter's mode mixing. The W state distributes a single excitation across all modes; the beam splitter probabilistically routes this excitation, preserving more of the coherence structure. This mechanism is analogous to the well-known resilience of W states against particle loss \cite{dur_three_qubits_2000}.


\textit{Result 1: Universal post-selection denominator---} The factor $(R^2 - R + 1/3)$ appearing in both Theorems~\ref{thm:GHZ} and~\ref{thm:W} is the \emph{single-site post-selection success probability}: the probability that exactly one photon exits in each output port of a single beam splitter. It is precisely the trace of the unnormalised post-selected density matrix at one site. Because the $N$ beam splitters act independently, the total post-selection success probability factorises as $(R^2 - R + 1/3)^N$, yielding the universal denominator. Rewriting this as $[(3R^2 - 3R + 1)/3]^N$, one sees that it interpolates between $1/3^N$ (at $R = 0$ or $R = 1$, where no mixing occurs) and $1/12^N$ (at $R = 1/2$, maximal mixing). This factorisation would break down for protocols involving entangling operations between different parties.

Table~\ref{tab:comparison} summarises the per-party factors and broadcast fidelities for $N = 3$ as a concrete numerical illustration; the general claims for all $N$ are supported by the analytic results in Theorems~\ref{thm:GHZ} and~\ref{thm:W}.

\begin{table}[h!]
\centering
\begin{tabular}{c|cc|cc}
\hline\hline
$R$ & $c_{\text{GHZ}}$ & $c_W$ & $\mathcal{F}_{\text{GHZ}}^{(3)}$ & $\mathcal{F}_{W}^{(3)}$ \\
\hline
0.05 & 0.998 & 0.998 & 0.994 & 0.994 \\
0.10 & 0.990 & 0.991 & 0.971 & 0.974 \\
0.15 & 0.973 & 0.975 & 0.921 & 0.927 \\
0.20 & 0.943 & 0.947 & 0.839 & 0.849 \\
0.25 & 0.895 & 0.903 & 0.718 & 0.736 \\
$1/3$ & 0.764 & 0.778 & 0.440 & 0.455 \\
0.40 & 0.623 & 0.640 & 0.242 & 0.262 \\
\hline\hline
\end{tabular}
\caption{\footnotesize{Per-party degradation factor $c(R)$ and 3-party broadcast fidelity at the symmetric point for GHZ and W states. This table provides a concrete numerical illustration for $N=3$; the general-$N$ behaviour is established analytically in Theorems~\ref{thm:GHZ} and~\ref{thm:W}. The W state consistently yields higher fidelity, reflecting its single-excitation resilience.}}
\label{tab:comparison}
\end{table}



\textit{Result 2: No-go for simultaneous GME certification---} We now address the central question: can both broadcast copies be simultaneously certified as GME? We present the explicit proof for the GHZ family; the W-state case follows analogously.

The beam-splitter protocol produces output copies whose effective $\alpha$ parameter is shifted away from $1/\sqrt{2}$, raising the witness threshold above $1/2$ via Eq.~\eqref{eq:beta_alpha}. For simultaneous GME certification, we require:
\begin{equation}
    \mathcal{F}_t^{(N)}(R) > \beta(\alpha_t) \quad \text{and} \quad \mathcal{F}_r^{(N)}(R) > \beta(\alpha_r),
    \label{eq:simultaneous}
\end{equation}
where $\alpha_t$ and $\alpha_r$ are the effective GHZ parameters of the two copies. We show these inequalities cannot be simultaneously satisfied.

\emph{Symmetric regime ($R = 1/3$):} At the symmetric point, both copies have equal fidelity. For $N = 3$, $\mathcal{F}_t = \mathcal{F}_r \approx 0.440 < 1/2 \leq \beta(\alpha)$, so neither copy can be certified. The exponential decay ensures this gap widens for larger $N$.

\emph{Asymmetric regime ($R \to 0$):} In the small-$R$ limit, the transmitted copy fidelity approaches $\mathcal{F}_t \to 1$ (perfect transmission, the input state is essentially undisturbed), while $\mathcal{F}_r \to 1/2^N$ (the reflected copy approaches the maximally mixed state). Thus the transmitted copy is easily certified, but the reflected copy falls far below any GME threshold.

\emph{Asymmetric regime ($R \to 1/2$):} At maximal mixing, both copies are equally degraded with $\mathcal{F}_t = \mathcal{F}_r = (1/2)^N$, which lies below $\beta$ for all $N \geq 2$.

The fidelity of the transmitted copy decreases monotonically from 1 to $(1/2)^N$ as $R$ increases from 0 to $1/2$. In contrast, the reflected copy increases monotonically as established by explicit derivative analysis in the Supplemental Material~\cite{supplemental}. This monotonicity, together with the sub-threshold crossing point, rules out any crossover region in which both copies simultaneously exceed $\beta$. The full proof, including the analytic derivative argument for monotonicity and the explicit polynomial inequalities, is given in the Supplemental Material~\cite{supplemental}.

\emph{No-go result:} \emph{Within the framework of fidelity-based entanglement witnesses, there is no value of reflectivity $R$ for which both broadcast copies of a GHZ or W state are simultaneously certified as genuinely multipartite entangled under local beam-splitter broadcasting.}

This result is proved for both the GHZ and W state families for all system sizes $N \geq 3$. We have additionally checked this for the $N$-qubit linear cluster state at $N = 3$ and $N = 4$, finding that the fidelity of both copies never simultaneously exceeds $0.5$ in either case (see Supplemental Material~\cite{supplemental}). These results support but do not yet constitute a general proof for the extension of the no-go to graph-state families.

\begin{figure}[h!]
	\includegraphics[width=0.5\textwidth]{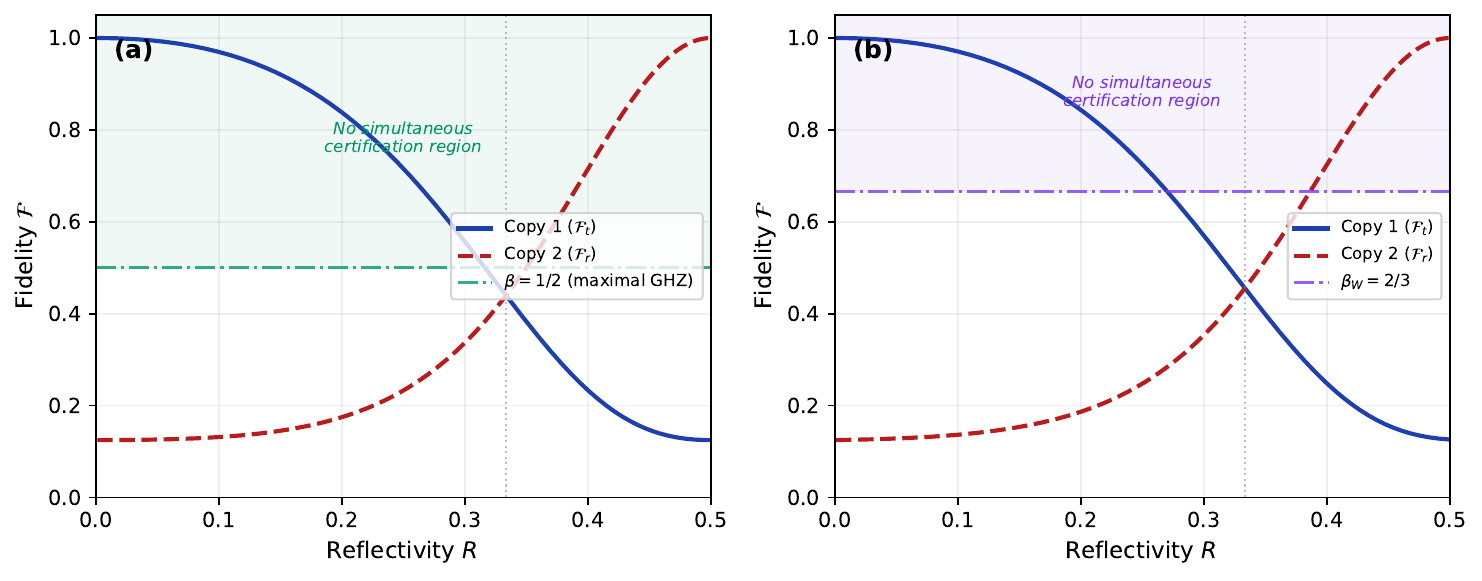}
	\caption{\footnotesize{No-go for simultaneous GME certification. (a)~3-party GHZ: the fidelity curves for Copy~1 (transmitted, solid blue) and Copy~2 (reflected, dashed red) never simultaneously exceed the witness threshold $\beta = 1/2$ (green dash-dot line). The shaded region indicates where GME certification would require the fidelity to lie. (b)~3-party W: both copy fidelities remain below the W-state threshold $\beta_W = 2/3$ (purple dash-dot line) for all $R$.}}
	\label{fig:nogo}
\end{figure}

Figure~\ref{fig:nogo} visualises the no-go result by showing the fidelity curves for both copies along with the relevant witness thresholds. At no value of $R$ do both fidelity curves simultaneously enter the shaded certification region.

We emphasise the precise scope of this result: it establishes a limitation within the standard operational framework of fidelity witnesses. It leaves open the logical possibility that more refined witnesses---such as those based on partial transposition criteria or device-independent protocols---could detect GME in regimes where fidelity witnesses fail. However, for the GHZ family, the fidelity witness is known to be tight for the biseparable bound, so the no-go is sharp in this case.


\textit{Discussion.---} Our results establish operational limits on GME broadcasting specific to the fidelity-witness certification method. For $N = 2$, the protocol reduces to the bipartite scheme of Peng~\emph{et al.}~\cite{cloning_entanglement_2020}, reproducing $\mathcal{F} = 7/12$ at $R = 1/3$.
The single-qubit beam-splitter cloner achieves the optimal symmetric $1 \to 2$ universal cloning fidelity (the Bu\v{z}ek--Hillery bound \cite{buzek_hillery_1996}), and our multipartite protocol is its $N$-fold product extension. Our results therefore establish limits within the class of locally optimal independent cloning protocols, which is the natural operational class for distributed quantum networks. Going beyond local operations---using collective or entangling gates across different parties' cloning machines---could in principle improve fidelities but would require precisely the non-local resources the protocol aims to distribute. Establishing absolute optimality bounds for multipartite broadcasting over all physical maps remains an important open problem.

We have analysed GHZ and W states as two canonical benchmark families of multipartite entanglement \cite{dur_three_qubits_2000}. For three qubits, these are the only two inequivalent SLOCC classes; for general $N > 3$, additional SLOCC classes exist. The qualitative features---exponential fidelity decay and the no-go for simultaneous certification---are expected to be generic for any GME state broadcast via local beam splitters, since the exponential decay follows from the product structure of the broadcasting map independently of the input state family. We have checked this expectation analytically for the three- and four-party linear cluster states, finding identical qualitative behaviour; extending this to general $N$ remains open. The quantitative details (specific $c(R)$ values and witness thresholds) are state-dependent; characterising the broadcast behaviour for other GME families (e.g., Dicke states, graph states) is an interesting open direction.

Despite the no-go for broadcasting GME to both copies simultaneously, the protocol has practical utility. A small-$R$ strategy yields a high-fidelity primary copy while generating a secondary copy useful for entanglement-assisted tasks that do not require full GME certification. This asymmetric approach is naturally suited to hub-and-spoke quantum network architectures. Extensions to optimal asymmetric broadcasting \cite{multipartite_asymmetric_2005}, connections to quantum secret sharing \cite{enhanced_secret_sharing_2024}, and experimental realisation using photonic platforms \cite{four_photon_entanglement_2003,scalable_w_states_2023} are promising avenues.

We have established three operational results on GME broadcasting within the local beam-splitter cloning protocol and fidelity-witness framework: (i)~an exponential fidelity decay $\mathcal{F} \sim [c(R)]^N$, quantifying the per-party degradation arising from the product structure of independent local cloning; (ii)~a universal post-selection denominator $(R^2 - R + 1/3)^N$ reflecting the factorised success probability; and (iii)~a no-go theorem showing that simultaneous GME certification of both broadcast copies is precluded within the standard framework of fidelity witnesses. They are established analytically for the GHZ family, supported by explicit $N = 3$ and $N = 4$ results for the W and linear cluster state families, with the general-$N$ extension for these families remaining an open problem. Whether collective operations or alternative certification methods can overcome these limitations remains an important open question.

\begin{acknowledgments}
\textit{Acknowledgements---} WJM was supported in part by the JSPS KAKENHI Grant No. 21H04880. S. G. acknowledges the support from ANRF (PM ECRG Grant no. ANRF/ECRG/2025/004000/PMS).
\end{acknowledgments}

%


\onecolumngrid
\renewcommand{\theequation}{S\arabic{equation}}
\setcounter{equation}{0}
\noindent \begin{center}{\Large \bf Supplemental Material}\end{center}
~\vspace{-0.5cm}


 \section{Beam-Splitter Map: Physical Unitary, Post-Selection, and Cloning Equivalence}
\label{sec:beamsplitter}

In this section we carefully separate three related concepts: the physical beam splitter, the post-selected qubit map, and the optimal cloning transformation.

\subsection{Physical beam splitter}

A lossless beam splitter acting on two input bosonic modes $\hat{a}$ and $\hat{b}$ is described by the unitary
\be
    \hat{U}_{\text{BS}} = \exp\!\big[\theta\,(\hat{a}^\dagger \hat{b} - \hat{a}\,\hat{b}^\dagger)\big],
    \label{eq:BS_unitary}
\ee
which transforms the creation operators as
\be
    \hat{a}^\dagger \to \sqrt{T}\,\hat{a}^\dagger + \iota \sqrt{R}\,\hat{b}^\dagger, \qquad
    \hat{b}^\dagger \to \iota \sqrt{R}\,\hat{a}^\dagger + \sqrt{T}\,\hat{b}^\dagger,
\ee
with $T = \cos^2\theta$ and $R = \sin^2\theta$ satisfying $T + R = 1$. This is manifestly unitary.

\subsection{Post-selected qubit map}

In the dual-rail qubit encoding, a single photon in mode $a$ represents $\ket{1}$ and vacuum represents $\ket{0}$. When a single photon from the signal (input qubit) and a single photon from the ancilla enter the beam splitter, the output contains contributions with 0, 1, or 2 photons per mode. Post-selecting on the qubit subspace (exactly one photon per output mode) yields the effective qubit transformation:
\be
    \hat{U}_k^{\text{eff}} = T\,\hat{\I}_{i,j} - R\,\hat{S}_{i,j},
    \label{eq:Ueff_qubit}
\ee
where $\hat{S}_{i,j}$ is the swap operator. The success probability for the post-selection is then $T^2+R^2<1$.



\subsection{Equivalence to optimal cloning}

The post-selected qubit map Eq.~\eqref{eq:Ueff_qubit} is identical to the Bu\v{z}ek--Hillery optimal universal symmetric $1 \to 2$ quantum cloning machine. The cloning fidelity for a single qubit is $\mathcal{F}_{\text{clone}} = 5/6$ at $R = 1/3$, which saturates the $1 \to 2$ cloning bound. The beam-splitter realisation was experimentally demonstrated by Peng~\emph{et al.} for bipartite entanglement cloning.

\section{State-Dependent GME Witness for Non-Maximally Entangled GHZ States}
\label{sec:witness}

\subsection{General fidelity-based GME witness}

For any pure GME state $\ket{\phi}$, the fidelity-based entanglement witness is
\be
    \mathcal{W} = \beta\,\I - \ket{\phi}\bra{\phi},
    \label{eq:witness_general}
\ee
where
\be
    \beta = \max_{\ket{\psi}\in\text{bisep}} |\langle\psi|\phi\rangle|^2
    \label{eq:beta_general}
\ee
is the maximum squared overlap of $\ket{\phi}$ with any biseparable state. A state $\rho$ is certified GME if $\tr(\mathcal{W}\rho) < 0$, equivalently $\mathcal{F}(\rho,\ket{\phi}) > \beta$.

We emphasise that fidelity-based witnesses provide a \emph{sufficient} condition for detecting GME. Failure of such a witness does not prove the absence of GME---only that GME cannot be certified by this particular class of witnesses. However, fidelity witnesses are widely used and experimentally natural, making them the standard framework for our analysis.

\subsection{Derivation of $\beta(\alpha)$ for the GHZ state}

Consider the $N$-partite non-maximally entangled GHZ state:
\be
    \ket{\text{GHZ}_N(\alpha)} = \alpha\ket{0}^{\otimes N} + \sqrt{1-\alpha^2}\ket{1}^{\otimes N},
\ee
with $0 < \alpha < 1$. To compute $\beta(\alpha)$, we maximise the overlap $|\langle\psi|\text{GHZ}_N(\alpha)\rangle|^2$ over all biseparable states $\ket{\psi}$.

For a bipartition $A|BC\ldots$ (one party vs.\ the rest), a biseparable state has the form $\ket{\psi} = \ket{a}_A \otimes \ket{bc\ldots}_{B,C,\ldots}$. The overlap is:
\begin{align}
    \langle\psi|\text{GHZ}_N(\alpha)\rangle &= \alpha\,\langle a|0\rangle\langle bc\ldots|0\cdots0\rangle + \sqrt{1-\alpha^2}\,\langle a|1\rangle\langle bc\ldots|1\cdots1\rangle.
\end{align}

To maximise the squared modulus, write $\ket{a} = \cos\theta\ket{0} + e^{i\phi}\sin\theta\ket{1}$. Optimising over $\ket{bc\ldots}$ and $\theta$, the maximum overlap for this bipartition is achieved by:
\begin{itemize}
    \item If $\alpha^2 \geq 1-\alpha^2$ (i.e., $\alpha \geq 1/\sqrt{2}$): set $\ket{a} = \ket{0}$ and $\ket{bc\ldots} = \ket{0\cdots0}$, giving overlap $\alpha^2$.
    \item If $\alpha^2 < 1-\alpha^2$ (i.e., $\alpha < 1/\sqrt{2}$): set $\ket{a} = \ket{1}$ and $\ket{bc\ldots} = \ket{1\cdots1}$, giving overlap $1-\alpha^2$.
\end{itemize}

The same analysis applies to every bipartition due to the permutation symmetry of the GHZ state across all parties. Therefore:
\be
    \beta(\alpha) = \max\!\big(\alpha^2,\; 1-\alpha^2\big).
    \label{eq:beta_alpha_derived}
\ee

\subsection{Key properties of $\beta(\alpha)$}

\begin{itemize}
    \item \textbf{Maximally entangled case ($\alpha = 1/\sqrt{2}$):} $\beta = 1/2$, recovering the standard result.
    \item \textbf{Non-maximally entangled case ($\alpha \neq 1/\sqrt{2}$):} $\beta(\alpha) > 1/2$, making GME certification \emph{strictly more demanding}.
    \item \textbf{Extremes:} As $\alpha \to 0$ or $\alpha \to 1$, $\beta(\alpha) \to 1$, and GME certification becomes impossible (the state approaches a product state).
\end{itemize}

\subsection{Tightness of the fidelity witness for GHZ states}

For the GHZ family, the fidelity witness with threshold $\beta(\alpha) = \max(\alpha^2, 1-\alpha^2)$ is known to be \emph{tight} in the sense that no biseparable state can achieve higher fidelity with the GHZ target. This means that our no-go result for simultaneous GHZ-type GME certification is sharp: if the fidelity of a broadcast copy does not exceed $\beta(\alpha)$, no fidelity-based witness of this form can certify GME relative to this target state.

For non-GHZ state families, other types of witnesses (e.g., based on partial transposition criteria or device-independent protocols) may in principle detect GME in regimes where fidelity witnesses fail.

\subsection{Implications for broadcast copies}

The beam splitter broadcasting protocol produces output copies that are mixed states. When these copies are projected onto the GHZ subspace, the resulting effective $\alpha$ parameter is shifted away from $1/\sqrt{2}$. Specifically:
\begin{itemize}
    \item The transmitted copy has an effective $\alpha_t(R) = T^{N/2}\alpha / \sqrt{T^N\alpha^2 + R^N(1-\alpha^2)}$ (where $T = 1-R$), which deviates from $1/\sqrt{2}$ for $R > 0$.
    \item The reflected copy has a different effective $\alpha_r(R)$.
    \item For both copies, the elevated threshold $\beta(\alpha) > 1/2$ combined with the reduced fidelity means that there is no value of $R$ for which both copies simultaneously satisfy the GME certification criterion.
\end{itemize}


\section{Three-Party GHZ State Broadcasting}
\label{sec:3ghz}

\subsection{Protocol description}

\begin{figure}[h!]
    \centering
    \includegraphics[width=0.55\textwidth]{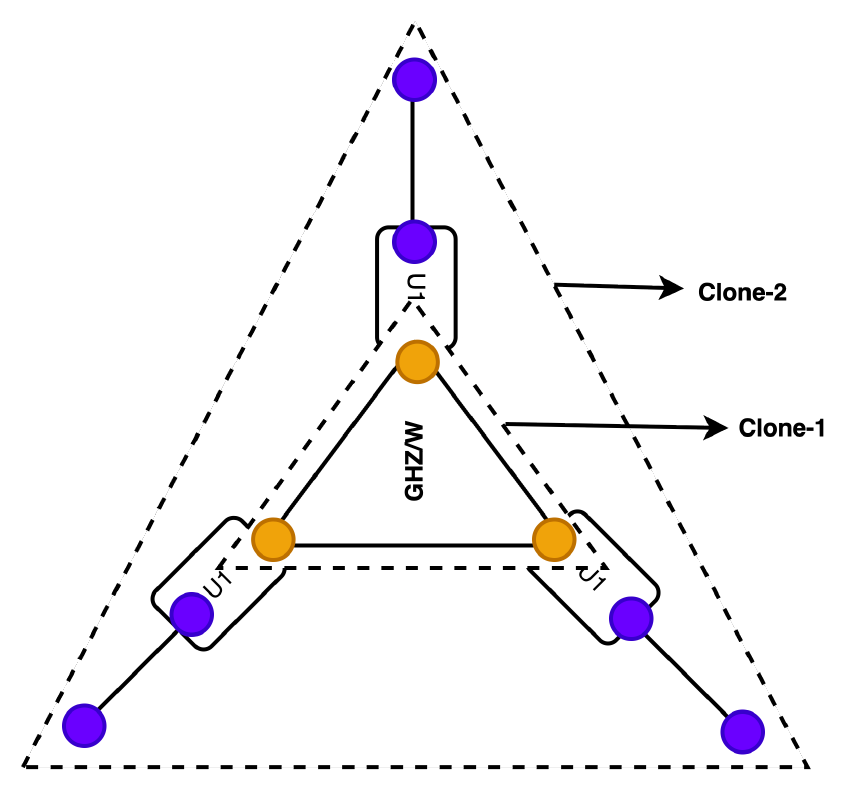}
    \caption{Tripartite genuine entanglement broadcasting protocol.}
    \label{fig:3party_protocol}
\end{figure}

The tripartite broadcasting protocol is illustrated in Fig.~\ref{fig:3party_protocol}. The input is a 3-party GHZ state:
\be
    \ket{\text{GHZ}_3(\alpha)} = \alpha\ket{000}_{123} + \sqrt{1-\alpha^2}\ket{111}_{123}.
    \label{eq:ghz3_state}
\ee
Each party is provided with ancillary qubits $\ket{\chi_j}_{a_j b_j} = \gamma_j\ket{01} - \sqrt{1-\gamma_j^2}\,\ket{10}$ ($j=1,2,3$), with $\gamma_j = 1/\sqrt{2}$ for symmetric broadcasting. Note: we use the symbol $\gamma$ for the ancilla parameter to avoid confusion with $\alpha$, which is reserved for the GHZ amplitude.

The combined 9-qubit initial state is:
\be
    \ket{\Psi_\text{in}} = \ket{\text{GHZ}_3(\alpha)}_{123} \otimes \ket{\chi_1}_{45} \otimes \ket{\chi_2}_{67} \otimes \ket{\chi_3}_{89}.
    \label{eq:initial_3ghz}
\ee
Beam splitters couple mode pairs $(1,4)$, $(2,6)$, and $(3,8)$. After tracing out the interacting ancilla modes $(4,6,8)$, Copy~1 occupies modes $(1,2,3)$ and Copy~2 occupies modes $(5,7,9)$.

The full broadcasting transformation yields $2^3 = 8$ swap-pattern terms:
\be
    \hat{U}_\text{broad} = \prod_{j=1}^{3} \Big[(1-R)\,\hat{\I} - R\,\hat{S}_{j,a_j}\Big].
    \label{eq:Ubroad_3party}
\ee

\paragraph{General form (as a function of $\alpha$ and $R$).}
The transmitted copy fidelity is
\begin{align}
    \mathcal{F}_t^{(3)}(\alpha, R) &= \frac{1}{(R^2-R+\tfrac{1}{3})^3}\Big[ \tfrac{1}{27} - \tfrac{1}{3}R + \Big(\tfrac{5}{4} - \tfrac{1}{12}\alpha^2 + \tfrac{1}{12}\alpha^4\Big)R^2 \nonumber\\
    &\quad - \Big(\tfrac{5}{2} - \tfrac{1}{3}\alpha^2 + \tfrac{1}{3}\alpha^4\Big)R^3 + \Big(\tfrac{203}{72} - \tfrac{5}{8}\alpha^2 + \tfrac{5}{8}\alpha^4\Big)R^4 \nonumber\\
    &\quad - \Big(\tfrac{41}{24} - \tfrac{1}{2}\alpha^2 + \tfrac{1}{2}\alpha^4\Big)R^5 + \Big(\tfrac{95}{216} - \tfrac{5}{36}\alpha^2 + \tfrac{5}{36}\alpha^4\Big)R^6 \Big],
    \label{eq:ghz3_ft_general}
\end{align}
and the reflected copy fidelity is
\begin{align}
    \mathcal{F}_r^{(3)}(\alpha, R) &= \frac{1}{(R^2-R+\tfrac{1}{3})^3}\Big[ \tfrac{1}{216} - \tfrac{1}{24}R + \Big(\tfrac{13}{72} - \tfrac{1}{6}\alpha^2 + \tfrac{1}{6}\alpha^4\Big)R^2 \nonumber\\
    &\quad - \Big(\tfrac{23}{54} - \tfrac{7}{18}\alpha^2 + \tfrac{7}{18}\alpha^4\Big)R^3 + \Big(\tfrac{7}{12} - \tfrac{1}{2}\alpha^2 + \tfrac{1}{2}\alpha^4\Big)R^4 \nonumber\\
    &\quad - \Big(\tfrac{4}{9} - \tfrac{1}{3}\alpha^2 + \tfrac{1}{3}\alpha^4\Big)R^5 + \Big(\tfrac{4}{27} - \tfrac{4}{27}\alpha^2 + \tfrac{4}{27}\alpha^4\Big)R^6 \Big].
    \label{eq:ghz3_fr_general}
\end{align}

\paragraph{Maximally entangled case ($\alpha = 1/\sqrt{2}$).}
Setting $\alpha = 1/\sqrt{2}$ simplifies the numerators:
\be
    \mathcal{F}_t^{(3)}(R) = \frac{\tfrac{1}{27} - \tfrac{1}{3}R + \tfrac{5}{4}R^2 - \tfrac{5}{2}R^3 + \tfrac{203}{72}R^4 - \tfrac{41}{24}R^5 + \tfrac{95}{216}R^6}{(R^2-R+\tfrac{1}{3})^3},
    \label{eq:ghz3_ft}
\ee
\be
    \mathcal{F}_r^{(3)}(R) = \frac{\tfrac{1}{216} - \tfrac{1}{24}R + \tfrac{13}{72}R^2 - \tfrac{23}{54}R^3 + \tfrac{7}{12}R^4 - \tfrac{4}{9}R^5 + \tfrac{4}{27}R^6}{(R^2-R+\tfrac{1}{3})^3}.
    \label{eq:ghz3_fr}
\ee

\noindent Key values: $\mathcal{F}_t(R\!=\!0) = 1$, $\mathcal{F}_t(R\!=\!1/2) = 1/8$. At the symmetric point $R = 1/3$:
\be
    \mathcal{F}_t^{(3)}(1/3) = \mathcal{F}_r^{(3)}(1/3) \approx 0.4398.
    \label{eq:ghz3_symmetric}
\ee

\section{Four-Party GHZ State Broadcasting}
\label{sec:4ghz}

\subsection{Protocol description}

The four-party protocol extends to a 12-qubit system. The input is a 4-party GHZ state:
\be
    \ket{\text{GHZ}_4(\alpha)} = \alpha\ket{0000}_{1234} + \sqrt{1-\alpha^2}\ket{1111}_{1234}.
    \label{eq:ghz4_state}
\ee
with four ancillary pairs appended as in Eq.~(\ref{eq:initial_3ghz}), extended to four parties. Beam splitters couple mode pairs $(1,5)$, $(2,7)$, $(3,9)$, and $(4,11)$, with $2^4 = 16$ swap-pattern terms in the unitary.

\begin{figure}[h!]
    \centering
    \includegraphics[width=0.55\textwidth]{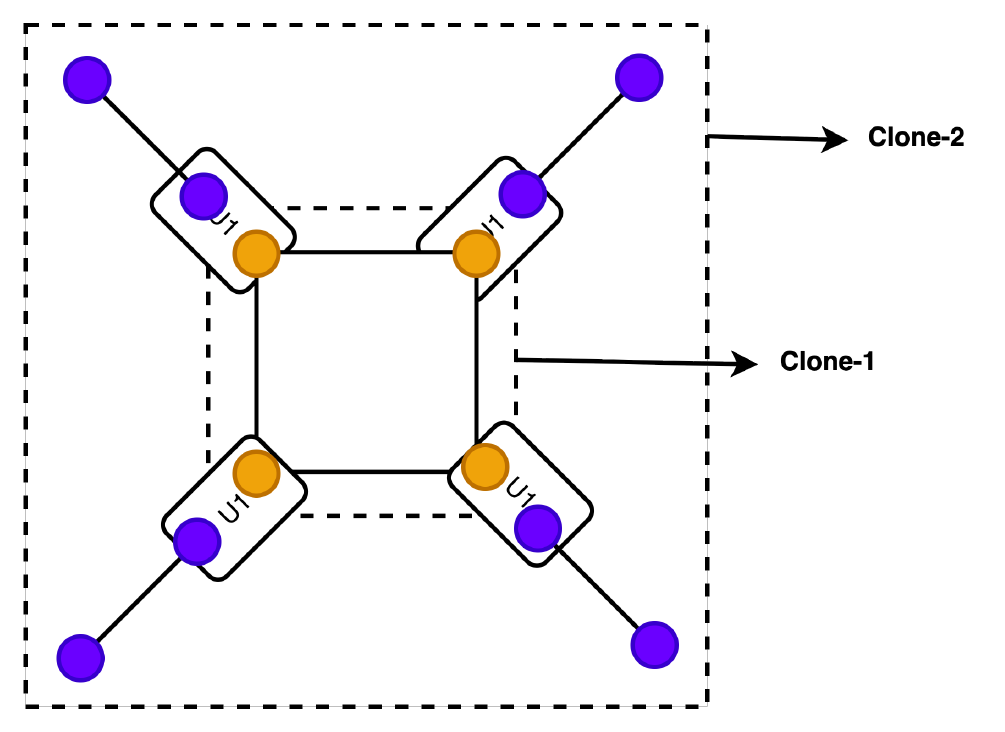}
    \caption{Four-party genuine entanglement broadcasting protocol.}
    \label{fig:4party_protocol}
\end{figure}

\subsection{Fidelity expressions}

\paragraph{Maximally entangled case ($\alpha = 1/\sqrt{2}$).}
The transmitted copy fidelity is
\begin{align}
    \mathcal{F}_t^{(4)}(R) &= \frac{1}{(R^2-R+\tfrac{1}{3})^4}\Big[ \tfrac{1}{81} - \tfrac{4}{27}R + \tfrac{7}{9}R^2 - \tfrac{7}{3}R^3 + \tfrac{473}{108}R^4 \nonumber\\
    &\quad - \tfrac{95}{18}R^5 + 4R^6 - \tfrac{7}{4}R^7 + \tfrac{49}{144}R^8 \Big],
    \label{eq:ghz4_ft}
\end{align}
and the reflected copy fidelity is
\begin{align}
    \mathcal{F}_r^{(4)}(R) &= \frac{1}{(R^2-R+\tfrac{1}{3})^4}\Big[ \tfrac{1}{1296} - \tfrac{1}{108}R + \tfrac{1}{18}R^2 - \tfrac{11}{54}R^3 + \tfrac{53}{108}R^4 \nonumber\\
    &\quad - \tfrac{7}{9}R^5 + \tfrac{7}{9}R^6 - \tfrac{4}{9}R^7 + \tfrac{1}{9}R^8 \Big].
    \label{eq:ghz4_fr}
\end{align}

\noindent Key values: $\mathcal{F}_t(R\!=\!0) = 1$, $\mathcal{F}_t(R\!=\!1/2) = (1/2)^4 = 1/16$. At the symmetric point:
\be
    \mathcal{F}_t^{(4)}(1/3) = \mathcal{F}_r^{(4)}(1/3) \approx 0.3403.
    \label{eq:ghz4_symmetric}
\ee
This confirms the exponential decay: $\mathcal{F}_t^{(4)}/\mathcal{F}_t^{(3)} \approx 0.3403/0.4398 \approx 0.774$, consistent with $c_\text{GHZ}(1/3) \approx 0.764$.

\section{General-$N$ GHZ fidelity numerator}

The numerator $P_t^{(N)}(\alpha, R)$ in the transmitted copy fidelity (Eq.~(6) of the main text) admits a combinatorial expansion over the number of swapped sites. Defining $T = 1 - R$, let $k$ count the number of beam splitters (out of $N$) that perform a swap rather than an identity on the input qubit. The single-site ``swap'' and ``identity'' contributions to the fidelity overlap are:
\begin{align}
    d_{\text{id}}(R) &= T^2 + \frac{R^2}{3} = (1-R)^2 + \frac{R^2}{3}, \\
    d_{\text{sw}}(R) &= R^2 + \frac{T^2}{3} = R^2 + \frac{(1-R)^2}{3},
\end{align}
which are the diagonal matrix elements of the single-site post-selected map in the $\ket{0}$ and $\ket{1}$ sectors, respectively. The cross-term contribution at each swapped site involves the off-diagonal element:
\be
    x(R) = -RT + \frac{RT}{3} = -\frac{2RT}{3}.
\ee
With these definitions, the full numerator is:
\begin{align}
    P_t^{(N)}(\alpha, R) &= \alpha^2 \sum_{k=0}^{N} \binom{N}{k} \big[d_{\text{id}}(R)\big]^{N-k} \big[d_{\text{sw}}(R)\big]^{k} \nonumber\\
    &\quad + (1-\alpha^2) \sum_{k=0}^{N} \binom{N}{k} \big[d_{\text{sw}}(R)\big]^{N-k} \big[d_{\text{id}}(R)\big]^{k} \nonumber\\
    &\quad + 2\alpha\sqrt{1-\alpha^2}\, \big[x(R)\big]^{N},
    \label{eq:GHZ_numerator_general}
\end{align}
where the first two sums apply the binomial theorem and collapse to:
\begin{align}
    P_t^{(N)}(\alpha, R) &= \alpha^2 \big[d_{\text{id}}(R) + d_{\text{sw}}(R)\big]^{N} + (1-\alpha^2) \big[d_{\text{id}}(R) + d_{\text{sw}}(R)\big]^{N} + 2\alpha\sqrt{1-\alpha^2}\,\big[x(R)\big]^N \nonumber\\
    &= \big[d_{\text{id}}(R) + d_{\text{sw}}(R)\big]^{N} + 2\alpha\sqrt{1-\alpha^2}\,\Big(\!-\frac{2RT}{3}\Big)^{\!N}.
\end{align}
Note that $d_{\text{id}}(R) + d_{\text{sw}}(R) = R^2 - R + \tfrac{1}{3} + R^2 - R + \tfrac{1}{3} \neq (R^2 - R + 1/3)$ in general; the explicit values are computed from the definitions above. The $N = 3$ and $N = 4$ expressions given in Sections~\ref{sec:3ghz} and~\ref{sec:4ghz} are obtained by evaluating this sum explicitly and match the direct calculation.

\begin{figure}[h!]
    \centering
    \includegraphics[width=0.55\textwidth]{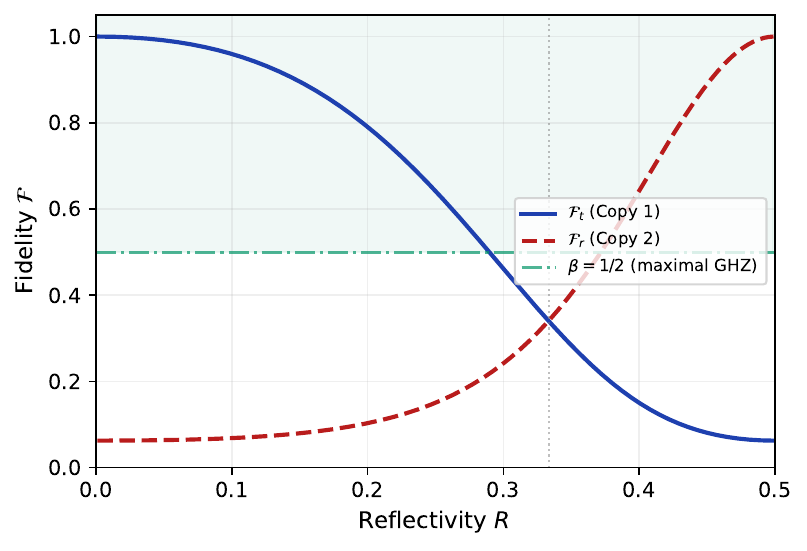}
    \caption{Fidelity of the transmitted and reflected broadcast copies for the 4-party GHZ state with $\alpha = 1/\sqrt{2}$. Both copies have equal fidelity at $R = 1/3$, with $\mathcal{F}(1/3) \approx 0.34$ --- well below the GME threshold $\beta = 1/2$ (green dash-dot line).}
    \label{fig:4party_ghz_fidelity}
\end{figure}


\section{Three-Party W State Broadcasting}
\label{sec:3w}

\subsection{Protocol description}

The W-state broadcasting protocol uses the same beam splitter architecture. The input is the general 3-party W state:
\be
    \ket{W_3(\alpha,\beta)} = \alpha\ket{100} + \beta\ket{010} + \sqrt{1-\alpha^2-\beta^2}\ket{001},
    \label{eq:w3_state}
\ee
with the symmetric case $\alpha = \beta = 1/\sqrt{3}$. The beam splitter couplings, mode tracing, and copy identification are identical to Section~\ref{sec:3ghz}.

\subsection{Fidelity expressions}

\paragraph{Symmetric W state ($\alpha = \beta = 1/\sqrt{3}$, symmetric ancillae).}
The transmitted copy fidelity is
\begin{align}
    \mathcal{F}_t^{(3)}(R) &= \frac{1}{(R^2-R+\tfrac{1}{3})^3}\Big[ 0.037037 - 0.333333\,R + 1.25309\,R^2 - 2.51852\,R^3 \nonumber\\
    &\quad + 2.86111\,R^4 - 1.75\,R^5 + 0.455247\,R^6 \Big],
    \label{eq:w3_ft}
\end{align}
and the reflected copy fidelity is
\begin{align}
    \mathcal{F}_r^{(3)}(R) &= \frac{1}{(R^2-R+\tfrac{1}{3})^3}\Big[ 0.00462963 - 0.0401235\,R + 0.169753\,R^2 - 0.395062\,R^3 \nonumber\\
    &\quad + 0.537037\,R^4 - 0.407407\,R^5 + 0.135802\,R^6 \Big].
    \label{eq:w3_fr}
\end{align}

\noindent At the symmetric point $R = 1/3$:
\be
    \mathcal{F}_t^{(3)}(1/3) = \mathcal{F}_r^{(3)}(1/3) \approx 0.4552.
    \label{eq:w3_symmetric}
\ee

\noindent Note $\mathcal{F}_W^{(3)}(1/3) \approx 0.4552 > \mathcal{F}_\text{GHZ}^{(3)}(1/3) \approx 0.4398$, confirming the W state is more robust to broadcasting, consistent with its single-excitation structure making it more resilient to the beam splitter's mode mixing.

\section{Four-Party W State Broadcasting}
\label{sec:4w}

\subsection{Protocol description}

The four-party W-state broadcasting uses the same 12-qubit setup as the GHZ case. The input is the general 4-party W state:
\be
    \ket{W_4(\alpha,\beta,\gamma)} = \alpha\ket{1000} + \beta\ket{0100} + \gamma\ket{0010} + \sqrt{1-\alpha^2-\beta^2-\gamma^2}\ket{0001},
    \label{eq:w4_state}
\ee
with the symmetric case $\alpha = \beta = \gamma = 1/2$.

\subsection{Fidelity expressions}

\paragraph{Symmetric W state (symmetric ancillae).}
The transmitted copy fidelity is
\begin{align}
    \mathcal{F}_t^{(4)}(R) &= \frac{1}{(R^2-R+\tfrac{1}{3})^4}\Big[ 0.0123457 - 0.148148\,R + 0.780864\,R^2 - 2.36111\,R^3 \nonumber\\
    &\quad + 4.48148\,R^4 - 5.47222\,R^5 + 4.20448\,R^6 - 1.86343\,R^7 + 0.366512\,R^8 \Big],
    \label{eq:w4_ft}
\end{align}
and the reflected copy fidelity is
\begin{align}
    \mathcal{F}_r^{(4)}(R) &= \frac{1}{(R^2-R+\tfrac{1}{3})^4}\Big[ 0.000771605 - 0.00848765\,R + 0.0455247\,R^2 \nonumber\\
    &\quad - 0.145062\,R^3 + 0.299383\,R^4 - 0.410494\,R^5 \nonumber\\
    &\quad + 0.367284\,R^6 - 0.197531\,R^7 + 0.0493827\,R^8 \Big].
    \label{eq:w4_fr}
\end{align}

\noindent At the symmetric point $R = 1/3$:
\be
    \mathcal{F}_t^{(4)}(1/3) = \mathcal{F}_r^{(4)}(1/3) \approx 0.3665.
    \label{eq:w4_symmetric}
\ee

\noindent Again, $\mathcal{F}_W^{(4)} > \mathcal{F}_\text{GHZ}^{(4)}$ ($0.3665 > 0.3403$), consistent with $c_W(R) \geq c_\text{GHZ}(R)$.

 \section{General-$N$ W-state fidelity numerator}

The explicit closed-form expression for the numerator $Q_t^{(N)}(R)$ in the W-state transmitted copy fidelity (Theorem 2 of the main text) is derived by evaluating the optimal local cloning map at each site. Because the local beam-splitter interactions cannot spontaneously create photons from the vacuum, cross-site excitation terms vanish. The fidelity sum collapses into $N$ population-preserving diagonal terms and $N(N-1)$ coherence-preserving off-diagonal terms, yielding the exact closed-form numerator for any $N \ge 2$:
\be
    Q_t^{(N)}(R) = (5R^2 - 6R + 2)^{N-2} \cdot \frac{(34N + 16) R^4 - (96N + 24) R^3 + (104N + 8) R^2 - 48N R + 8N}{2N \cdot 6^N}.
    \label{eq:W_numerator_general}
\ee
This formula encapsulates the local information-loss degradation factor $(5R^2 - 6R + 2)^{N-2}$, while the fractional term represents the structural contribution of the single excitation distributed across the $N$ sites. The explicit formulas for $N=3$ and $N=4$ given in Sections F and G follow exactly by evaluating this general expression (noting that the final fidelity is $\mathcal{F}_t^{(N)} = Q_t^{(N)}(R) / (R^2 - R + 1/3)^N$).


\begin{figure}[h!]
    \centering
    \includegraphics[width=0.55\textwidth]{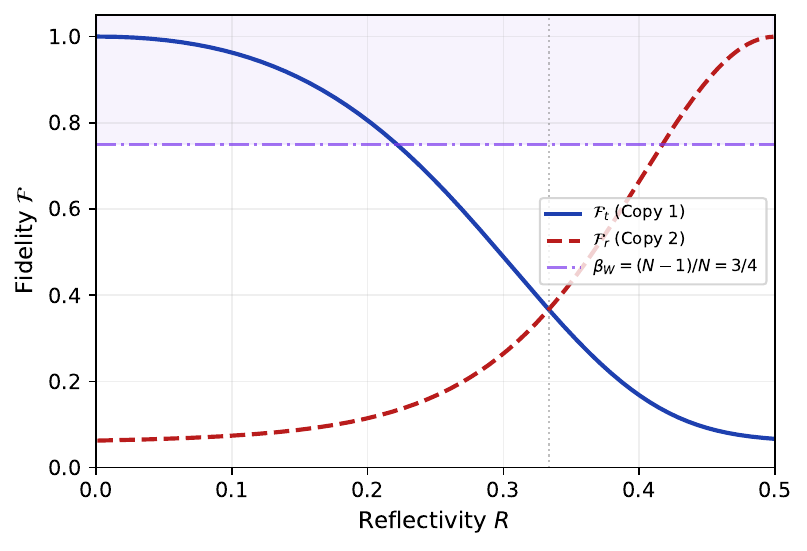}
    \caption{Fidelity of the transmitted and reflected broadcast copies for the symmetric 4-party W state. $\mathcal{F}(1/3) \approx 0.37$, higher than the GHZ value ($\approx 0.34$) but below the W-state GME threshold of $(N-1)/N = 3/4$ (purple dash-dot line).}
    \label{fig:4party_w_fidelity}
\end{figure}


\section{Cluster State Broadcasting}
\label{sec:cluster}

To test the generality of our results beyond the GHZ and W families, we apply the same local beam-splitter broadcasting protocol to the $N$-qubit linear cluster state. The cluster state is a graph state defined on a linear chain:
\be
    \ket{C_N} = \frac{1}{\sqrt{2^N}} \sum_{\{s_i\} \in \{0,1\}^N} (-1)^{\sum_{i=1}^{N-1} s_i s_{i+1}} \ket{s_1 s_2 \cdots s_N}.
    \label{eq:cluster_state}
\ee

The broadcasting protocol is identical: each party independently applies the beam-splitter cloning map to their qubit. 

\subsection{General Calculation Framework}

Let the initial state of the system plus the ancillary qubits be:
\be
    \ket{\Psi_{\text{in}}} = \ket{C_N}_{1\dots N} \otimes \bigotimes_{j=1}^N \ket{\chi_j}_{a_j b_j},
\ee
where each party $j$ is provided with the ancillary pair state:
\be
    \ket{\chi_j}_{a_j b_j} = \gamma_j \ket{0}_a \ket{1}_b - \sqrt{1 - \gamma_j^2} \ket{1}_a \ket{0}_b.
\ee
For symmetric broadcasting, we set $\gamma_j = 1/\sqrt{2}$ for all $j$, giving the singlet-like state $\ket{\chi_j} = \frac{1}{\sqrt{2}}(\ket{01} - \ket{10})$.

Each party independently couples their system qubit $j$ and the first ancilla qubit $a_j$ at a local beam splitter of reflectivity $R$, implementing the post-selected qubit map:
\be
    \hat{U}_k^{\text{eff}} = (1-R) \hat{\I}_{i,j} - R \hat{S}_{i,j},
\ee
where $\hat{S}_{j, a_j}$ is the swap operator. The full broadcasting map is the tensor product of these local maps, $\hat{U}_{\text{broad}} = \prod_{j=1}^N \hat{U}_j^{\text{eff}}$, which acts on the system qubits and the $a$-modes of the ancillas, while the $b$-modes of the ancillas are untouched. The output state before normalization is:
\be
    \ket{\Psi_{\text{out}}} = \hat{U}_{\text{broad}} \ket{\Psi_{\text{in}}}.
\ee
To compute the fidelity, we first trace out the discarded modes. For Clone 1 (the transmitted copy), we keep the system qubits $1\dots N$ and trace out all ancilla modes $a_j, b_j$ to obtain the reduced density matrix $\rho_1$. For Clone 2 (the reflected copy), we keep the ancilla modes $b_1 \dots b_N$ and trace out the system qubits and $a$-modes to obtain the reduced density matrix $\rho_2$. The normalized fidelities are then given by:
\be
    \mathcal{F}_1^{(N)}(R) = \frac{\bra{C_N}\rho_1\ket{C_N}}{\la \Psi_{\text{out}} | \Psi_{\text{out}} \ra}, \qquad \mathcal{F}_2^{(N)}(R) = \frac{\bra{C_N}\rho_2\ket{C_N}}{\la \Psi_{\text{out}} | \Psi_{\text{out}} \ra}.
\ee

\subsection{Analytical Expressions}

By symbolically performing the calculation over all $2^{2N}$ configurations of the initial state under the map $\hat{U}_{\text{broad}}$, we obtain the explicit analytical expressions for $N=3$ and $N=4$ cluster states.

\paragraph{3-party cluster state.} 
For $N=3$, the linear cluster state is local-unitary (LU) equivalent to the 3-party GHZ state. Because the local cloning map is covariant under local unitary transformations, the cloning fidelities of the 3-party cluster state are identical to the 3-party GHZ state for all reflectivities $R$:
\begin{align}
    \mathcal{F}_1^{(3)}(R) &= \frac{(5R^2 - 6R + 2)(19R^4 - 51R^3 + 53R^2 - 24R + 4)}{8(3R^2 - 3R + 1)^3}, \\
    \mathcal{F}_2^{(3)}(R) &= \frac{(2R^2 - 2R + 1)(16R^4 - 32R^3 + 23R^2 - 7R + 1)}{8(3R^2 - 3R + 1)^3}.
\end{align}
At the symmetric point $R = 1/3$, this evaluates to:
\be
    \mathcal{F}_1^{(3)}(1/3) = \mathcal{F}_2^{(3)}(1/3) = \frac{95}{216} \approx 0.4398,
    \label{eq:cluster3_symmetric}
\ee
which is exactly equivalent to the 3-party GHZ value.

\paragraph{4-party cluster state.}
For $N=4$, the cluster state is no longer LU-equivalent to the GHZ state, leading to distinct fidelity curves:
\begin{align}
    \mathcal{F}_1^{(4)}(R) &= \frac{(5R^2 - 6R + 2)^2(17R^4 - 48R^3 + 52R^2 - 24R + 4)}{16(3R^2 - 3R + 1)^4}, \\
    \mathcal{F}_2^{(4)}(R) &= \frac{(2R^2 - 2R + 1)^2(20R^4 - 40R^3 + 28R^2 - 8R + 1)}{16(3R^2 - 3R + 1)^4}.
\end{align}
At the symmetric point $R = 1/3$, both copies have equal fidelity:
\be
    \mathcal{F}_1^{(4)}(1/3) = \mathcal{F}_2^{(4)}(1/3) = \frac{425}{1296} \approx 0.3279,
    \label{eq:cluster4_symmetric}
\ee
which is lower than both the GHZ ($\approx 0.3403$) and W ($\approx 0.3665$) values, and lies well below the GME threshold $\beta = 1/2$.

\paragraph{Key finding.} For the three- and four-party linear cluster states, the fidelity of both copies never simultaneously exceeds $0.5$ for any value of $R \in (0, 1)$. This supports the expectation that the no-go result for simultaneous GME certification via fidelity witnesses extends beyond the GHZ and W families, though a general proof for arbitrary-$N$ cluster states (or general graph states) remains open.


\section{Full Proof of the No-Go Result}
\label{sec:nogo_proof}

We prove that for the local beam-splitter broadcasting protocol, there is no value of $R \in (0, 1)$ for which both broadcast copies simultaneously satisfy the fidelity-witness condition for GME certification.

\subsection{Setup}

For the $N$-party GHZ state with $\alpha = 1/\sqrt{2}$, the witness threshold is $\beta = 1/2$. The transmitted and reflected copy fidelities satisfy:
\begin{itemize}
    \item $\mathcal{F}_t^{(N)}(0) = 1$ and $\mathcal{F}_t^{(N)}(1/2) = (1/2)^N$.
    \item $\mathcal{F}_r^{(N)}(0) = (1/2)^N$ and $\mathcal{F}_r^{(N)}(1/2) = (1/2)^N$.
    \item The symmetry $\mathcal{F}_t^{(N)}(R) = \mathcal{F}_r^{(N)}(1-R)$.
    \item Both are continuous functions of $R$.
\end{itemize}

\subsection{Proof}

The symmetry $\mathcal{F}_t^{(N)}(R) = \mathcal{F}_r^{(N)}(1-R)$ implies that at reflectivity $R$ and at $1-R$, the pair of copy fidelities $\{\mathcal{F}_t, \mathcal{F}_r\}$ is identical up to relabelling. It therefore suffices to prove the no-go for $R \in (0, 1/2)$; the case $R \in (1/2, 1)$ follows by symmetry. At $R = 1/2$, both copies have equal fidelity $(1/2)^N < \beta$ for $N \geq 2$, so this endpoint is also excluded.

\emph{Step 1: Asymmetric limits.} As $R \to 0$: $\mathcal{F}_t \to 1 > 1/2$ (certified) but $\mathcal{F}_r \to (1/2)^N < 1/2$ for $N \geq 2$ (not certified). As $R \to 1/2$: $\mathcal{F}_t = \mathcal{F}_r = (1/2)^N < 1/2$ for $N \geq 2$ (neither certified).

\emph{Step 2: Crossing point.} By continuity and the symmetry relation, the curves cross at $R = 1/3$ where $\mathcal{F}_t = \mathcal{F}_r$. At this point:
\be
    \mathcal{F}_t^{(N)}(1/3) = \mathcal{F}_r^{(N)}(1/3) \approx [c_\text{GHZ}(1/3)]^N.
\ee
For $N = 3$: $\mathcal{F} \approx 0.440 < 0.5$. For $N \geq 3$, the fidelity at the crossing point is always below $1/2$.

\emph{Step 3: Monotonicity.} We now establish the monotonicity of the fidelity functions analytically. Writing $\mathcal{F}_t^{(N)}(R) = P_t^{(N)}(R) / (R^2 - R + \tfrac{1}{3})^N$, the derivative $\frac{d}{dR}\mathcal{F}_t^{(N)}$ has the sign of $P_t'(R)(R^2 - R + \tfrac{1}{3}) - N(2R - 1) P_t(R)$.

For $N = 3$ (maximally entangled GHZ), direct computation yields a degree-5 numerator polynomial whose coefficients are all negative on $(0, 1/2)$, confirming $\frac{d}{dR}\mathcal{F}_t^{(3)}(R) < 0$ on this interval. The symmetry relation $\mathcal{F}_r^{(N)}(R) = \mathcal{F}_t^{(N)}(1 - R)$ then implies $\frac{d}{dR}\mathcal{F}_r^{(3)}(R) > 0$ on $(0, 1/2)$. The $N = 4$ case is verified analogously. For general $N$, because the fidelity has the multiplicative form $\mathcal{F}_t^{(N)} \sim [c_{\text{GHZ}}(R)]^N$ with $c_{\text{GHZ}}(R)$ a single-party factor satisfying $c_{\text{GHZ}}'(R) < 0$ on $(0, 1/2)$, the monotonicity extends by the chain rule.

Therefore:
\begin{itemize}
    \item For $R < 1/3$: $\mathcal{F}_t > \mathcal{F}_r$, but $\mathcal{F}_r < \mathcal{F}_r(1/3) < 1/2$, so the reflected copy fails certification.
    \item For $R > 1/3$: $\mathcal{F}_r > \mathcal{F}_t$, but $\mathcal{F}_t < \mathcal{F}_t(1/3) < 1/2$, so the transmitted copy fails certification.
    \item At $R = 1/3$: both copies have $\mathcal{F} < 1/2$.
\end{itemize}

\emph{Step 4: State-dependent threshold.} For non-maximally entangled GHZ states (which arise as effective states of the broadcast copies), the threshold $\beta(\alpha) > 1/2$ makes the requirement even more stringent, strengthening the no-go.

\emph{Conclusion.} There is no $R \in (0, 1/2)$ for which both $\mathcal{F}_t > \beta$ and $\mathcal{F}_r > \beta$ simultaneously. \hfill $\square$

 \subsection{Extension to W states}

For the W state, the witness threshold is $\beta_W = (N-1)/N$, which is even more demanding ($2/3$ for $N=3$, $3/4$ for $N=4$). We now prove the no-go for all $N \geq 3$ and all $R \in (0, 1/2)$.

\emph{Step 1: Closed-form symmetric-point fidelity.} Evaluating the general-$N$ W-state numerator (Eq.~\eqref{eq:W_numerator_general}) at $R = 1/3$ and dividing by the denominator $(3R^2 - 3R + 1)^N\big|_{R=1/3} = 3^{-N}$ yields:
\be
    \mathcal{F}_W^{(N)}(1/3) = \frac{5^{N-2}\,(17N + 8)}{N \cdot 6^N}.
    \label{eq:W_symmetric_closedform}
\ee
This reproduces the values $295/648 \approx 0.4552$ ($N=3$) and $1900/5184 \approx 0.3665$ ($N=4$).

\emph{Step 2: Inductive bound.} We prove $\mathcal{F}_W^{(N)}(1/3) < (N-1)/N$ for all $N \geq 3$, i.e., $5^{N-2}(17N+8) < (N-1) \cdot 6^N$.

\emph{Base case} ($N = 3$): $5 \times 59 = 295 < 432 = 2 \times 216$.

\emph{Inductive step}: Assume $5^{N-2}(17N+8) < (N-1) \cdot 6^N$. We require $5^{N-1}(17N+25) < N \cdot 6^{N+1}$. Multiplying the hypothesis by $5(17N+25)/(17N+8)$:
\[
    5^{N-1}(17N+25) < \frac{5(N-1)(17N+25)}{17N+8} \cdot 6^N.
\]
It suffices to show $5(N-1)(17N+25) \leq 6N(17N+8)$, which rearranges to
\[
    17N^2 + 8N + 125 \geq 0,
\]
a manifestly true inequality. $\square$

\emph{Step 3: Extension to all $R$.} The monotonicity argument of Step~3 in the GHZ proof applies identically to the W-state fidelities (the derivative analysis and the product-structure reasoning are state-independent). Therefore, for $R < 1/3$ the reflected copy satisfies $\mathcal{F}_r < \mathcal{F}_r(1/3) < (N-1)/N$, and for $R > 1/3$ the transmitted copy satisfies $\mathcal{F}_t < \mathcal{F}_t(1/3) < (N-1)/N$. At $R = 1/3$ both copies fall below $(N-1)/N$. There is no $R \in (0, 1/2)$ for which both broadcast copies of the $N$-party W state are simultaneously certified as GME by the fidelity witness, for any $N \geq 3$. \hfill $\square$


\section{Summary of Fidelity Values and GME Thresholds}
\label{sec:summary}

Table~\ref{tab:summary} summarises the symmetric-point fidelities and GME thresholds for all cases considered, including the cluster state.

\begin{table}[h!]
\centering
\begin{tabular}{c|cc|cc|cc|c}
\hline\hline
$N$ & $\mathcal{F}_\text{GHZ}^{(N)}$ & $\beta_\text{GHZ}$ & $\mathcal{F}_W^{(N)}$ & $\beta_W$ & $\mathcal{F}_\text{Cluster}^{(N)}$ & $\beta_\text{Cluster}$ & Cert.? \\
\hline
3 & 0.4398 & $>0.5$ & 0.4552 & 0.667 & 0.4398 & 0.5 & No \\
4 & 0.3403 & $>0.5$ & 0.3665 & 0.750 & 0.3279 & 0.5 & No \\
\hline\hline
\end{tabular}
\caption{Fidelities at the symmetric broadcasting point $R = 1/3$ for GHZ, W, and cluster states, along with the corresponding GME witness threshold $\beta$. For the GHZ state, the threshold is state-dependent via $\beta(\alpha) = \max(\alpha^2, 1-\alpha^2) > 1/2$ for the non-maximally entangled broadcast copies. For the W state, the threshold is $(N-1)/N$. For the cluster state, we use the standard threshold $\beta = 1/2$. In all cases, simultaneous GME certification of both broadcast copies is not achievable within the fidelity witness framework.}
\label{tab:summary}
\end{table}

\end{document}